\documentclass[twocolumn,preprintnumbers,amsmath,amssymb,groupedaddress,superscriptaddress]{revtex4}
\usepackage{graphicx}
\usepackage{url}

\begin{document}
%
%
%

\title{Slow light in paraffin-coated Rb vapor cells}

\author{M.\ Klein${{}^{*}}{\dagger\ddagger}$\thanks{${}^{*}$Corresponding author. E-mail:mklein@fas.harvard.edu},
I.\ Novikova$\dagger$,
D.\ F.\ Phillips$\dagger$,
and R.\ L.\ Walsworth$\dagger\ddagger$\\
$\dagger$ Harvard-Smithsonian Center for Astrophysics, Cambridge, MA, 02138, USA\\
$\ddagger$  Department of Physics, Harvard University, Cambridge,MA, 02138, USA
\thanks{\vspace{6pt}\newline\centerline{\tiny{ {\em Journal of Modern Optics} ISSN 0950-0340 print/ ISSN 1362-3044 online
\textcopyright 2004 Taylor \& Francis Ltd}}
\newline\centerline{\tiny{ http://www.tandf.co.uk/journals}}\newline \centerline{\tiny{DOI:
10.1080/09500340xxxxxxxxxxxx}}}
}
 \received{\today}
%
%
%
%
%
\begin{abstract}
  We present preliminary results from an experimental study of slow
  light in anti-relaxation-coated Rb vapor cells, and describe the
  construction and testing of such cells. The slow ground state
  decoherence rate allowed by coated cell walls leads to a
  dual-structured electromagnetically induced transparency (EIT)
  spectrum with a very narrow ($<100$ Hz) transparency peak on top of
  a broad pedestal.  Such dual-structure EIT permits optical probe
  pulses to propagate with greatly reduced group velocity on two time
  scales.  We discuss ongoing efforts to optimize the pulse delay in such
  coated cell systems.
\end{abstract}
\maketitle

The manipulation of spin states in atomic ensembles lies at the heart
of many quantum-optical effects~\cite{scullybook,lukin03rmp}.
Such processes require high-quality state preparation and minimal
decoherence of the spin state of the atomic ensemble. In warm atomic
vapor cells, spin state lifetimes are often limited by wall
collisions, which thermalize internal atomic states and destroy spin
coherence. Coating the walls of the cell with a paraffin derivative
allows atoms to undergo many wall-collisions without losing their spin
coherence and prolongs spin lifetimes up to
$1$~second~\cite{bouchiat'66,robinson82,budker'98}.  Paraffin-coated
alkali-vapor cells have been successfully used to demonstrate spin
squeezing~\cite{kuzmichPRL00}, entanglement of atomic
ensembles~\cite{JulsgaardNature01} and quantum memory for continuous
quantum variables~\cite{schoriPRL02,JulsgaardNature04}, and are also
used for high-precision atomic clocks and
magnetometers~\cite{robinson83,aleksandrov'96,budker'00a}.

In this paper we study electromagnetically induced transparency
(EIT)~\cite{scullybook,harris'97pt,marangos'98} and slow
light~\cite{matsko_slreview,boyd_slreview} in paraffin-coated Rb
cells, \emph{e.g.,} as a step towards the application of such vapor
cells for a quantum memory~\cite{fleischhauer02pra,lukin03rmp} in a
quantum repeater~\cite{DLSZ} as well as for long delay times for
classical pulses.
In our present study, we characterize a dual-structured EIT spectrum
which includes a very narrow central feature that allows ultra-slow
group velocities for the propagation of weak classical pulses through
the EIT medium.


\section{Cell coating procedure \label{coating}}

\begin{figure*}
\includegraphics[width=1.6\columnwidth]{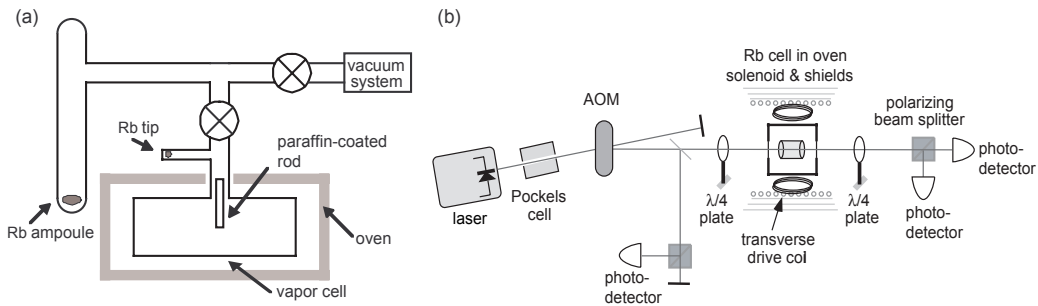}
\caption{(a) Schematic of the Pyrex manifold and vacuum system used to
  apply paraffin coating to the inside of the cell. (b) Schematic of
  the apparatus for double resonance, EIT, and slow light experiments.
  \emph{See text.} }
\label{schematics.fig}
\end{figure*}

Vapor cell coatings have been investigated for over 40 years utilizing
various techniques and derivatives of paraffin as a coating
material~\cite{robinson58, goldenberg61, bouchiat'66}.  Presently, we
employ a technique similar to Alley and co-workers~\cite{alley72}.
(In previous work~\cite{phillipsCoating}, we used techniques similar
to those described by Bouchiat~\cite{bouchiat'66} which was more
appropriate to spherical vapor cells.) Following
Robinson~\cite{robinson82}, we coat with tetracontane
(C${}_{40}$H${}_{82}$), a readily available component of paraffin.

After thoroughly cleaning the vapor cell manifold and attaching it to
the vacuum system (see Fig.~\ref{schematics.fig}a), we bake the
manifold under vacuum at $\gtrsim200\:{}^\circ$C.  This further cleans the
Pyrex glassware and reduces the base pressure of the system.
The manifold is then cooled to room temperature and filled with
N$_{2}$ gas to a pressure slightly above $1$ atmosphere. A valve on
the manifold is then removed allowing a glass rod, onto which a few
flakes of tetracontane have been melted, into the cell. The valve is
then reseated and the manifold evacuated.

To apply tetracontane to the cell walls, we enclose the cell in an
oven composed of a thin aluminum box with attached resistive heater
plates.  The neck of the cell manifold is wrapped in resistive heater
tape to provide additional heating of this region.  The valve is then
closed, isolating the cell from the rest of the vacuum system so that
tetracontane does not escape the coating region during the coating
process.  The cell and neck are then heated to $\sim200\:{}^\circ$C to
melt and vaporize the tetracontane with the neck region
$\gtrsim10\:{}^\circ$C warmer than the cell to keep tetracontane from
accumulating in the neck.  Finally, the cell is cooled to room
temperature and the valve opened to the vacuum system so that any gas
created during the coating process is pumped away.  Visible
tetracontane on the cell walls indicates cold spots during the coating
process and thus the lack of a uniform coating layer over the surface.
In this case, the process is repeated to improve the coating
uniformity.

After the coating is successfully applied, natural abundance metallic
Rb is distilled from its ampoule to the tip using a blown air heat
gun.  Bulk Rb is kept out of the main volume of the cell, as it can
interact with the coating and damage it.  Lastly, the coating rod is
moved out of the cell and the cell is ``pulled-off'' from the manifold
by melting the glass just above the Rb tip while also keeping the Rb
tip cool. Once the vapor cell has been completed, the temperature of
the Rb tip should be kept below that of the cell body so that bulk
rubidium remains in the tip rather than in the body. Additionally, the
cell body should be kept below $81\:{}^\circ$C, the melting point of
tetracontane.

\section{Apparatus\label{setup}}

The experimental apparatus shown in Fig.~\ref{schematics.fig}b is used
for testing the quality of the cell coating and for EIT and slow light
measurements. An external cavity diode laser~\cite{vortexLaser} tuned
near the Rb $\mathrm{D}_1$ line ($\lambda\approx795$~nm) produces
linearly polarized light which acts as a strong control field
($\Omega_C$ in Fig.~\ref{tests.fig}). If needed, a weak probe field
($\Omega_P$) is produced by rotating the optical polarization using a
Pockels cell with a maximum probe to control field intensity ratio of
10:1. The total intensity is regulated using an acousto-optic modulator
(AOM). After the AOM, a small fraction of the probe light is sent to a
photo-detector (PD) as a reference for slow light delay measurements.
Quarter-wave plates before and after the Rb cell convert the control
and probe field polarizations from linear to circular and back. The
maximum total laser power at the cell is $3$~mW, weakly focused into a
2 mm diameter beam, except where noted below.  The control and probe
components are then measured using a polarizing beam splitter and two
PDs.

The tetracontane-coated Rb cell is housed inside four layers of
magnetic shielding (to screen stray laboratory magnetic fields). The
cell is heated conductively by blowing hot air through the plastic
housing containing the vapor cell.
A solenoid and sets of coils mounted around the cell allow us to apply
a homogeneous magnetic field
and a transverse rf field ($\Omega_{\mathrm{rf}}$ in
Fig.~\ref{tests.fig}a) when needed.

\section{Coated cell testing~\label{test}}

\begin{figure*}
\includegraphics[width=1.6\columnwidth]{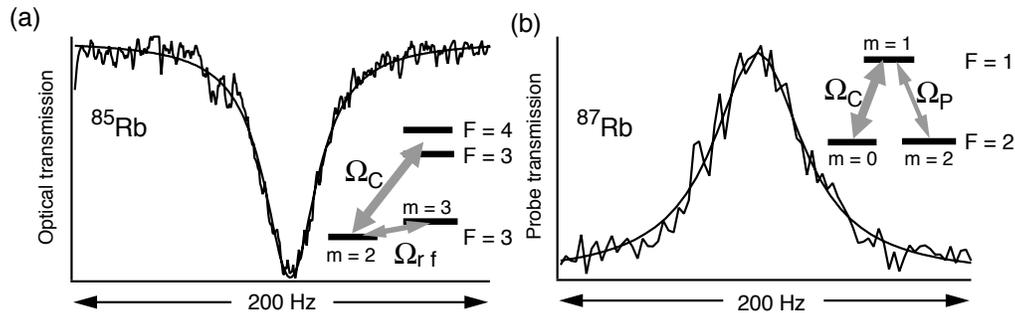}
\caption{Ground-state Zeeman coherence measurements using
  (a) optical pumping double resonance (FWHM $\approx22$~Hz) with
  $\approx 0.8$ mm beam diameter, and (b) EIT resonance (FWHM
  $\approx50$~Hz) with $\approx 2$ mm beam diameter. Smooth curves are
  Lorentzian fits to the data. Intensity $\lesssim0.1$ mW/cm${}^2$ to
  avoid power broadening.  \emph{Insets}: interaction schemes for each
  measurement.  Note that for the double resonance measurement we used
  ${}^{85}$Rb, which has a larger natural abundance ($72\%$) than
  ${}^{87}$Rb, providing higher Rb density at the same cell
  temperature.  }
\label{tests.fig}
\end{figure*}

We evaluate the coating quality by employing two complementary
techniques to measure the atomic spin decoherence rate.  In optical
pumping double resonance~\cite{demtroder_book}, we illuminate the cell
with circularly polarized light, optically pumping the atomic
population to the state with maximum angular momentum $m_F= F$, so
that the atomic vapor becomes transparent to the applied optical
field.  In the presence of a static longitudinal magnetic field which
splits the Zeeman ground state sublevels, we apply a transverse rf
magnetic field. If the rf field is resonant with the splitting of the
ground state Zeeman sublevels, it mixes the population into other
levels and thus increases the optical absorption.  Sweeping the rf
frequency through the Zeeman resonance produces a dip in the
transmission spectrum allowing a determination of both the Zeeman
frequency and its decoherence rate.
For the example results shown in Fig.~\ref{tests.fig}a, we detected
the change in transmission of a circularly polarized laser field tuned
to a transition between the $F=3$ ground state and unresolved excited
states of ${}^{85}$Rb in a static, longitudinal magnetic field of
$\approx38$~mG while sweeping the rf field frequency. The observed
transmission dip in Fig.~\ref{tests.fig}a had a full width of $22$~Hz,
corresponding to a Zeeman coherence lifetime $\approx 15$~ms.

Alternatively, we measure the EIT linewidth by applying continuous
control and probe optical fields with equal frequency and opposite
circular polarizations, which optically pump the atoms into a coherent
superposition of the ground-state Zeeman sublevels (\emph{i.e.,} a
``dark state'')~\cite{scullybook}.  Ideally, atoms in this state are
completely decoupled from both optical fields and do not absorb any
light. An applied magnetic field lifts the degeneracy of Zeeman
sublevels and destroys the dark state, leading to greater optical
absorption.
For the example results shown in Fig.~\ref{tests.fig}b (same Rb cell
as in Fig.~\ref{tests.fig}a), the control and probe fields were
resonant with the $F=2\rightarrow{}F^\prime=1$ transition of
${}^{87}$Rb, forming a $\Lambda$-system on the $m_{F}=0$ and
$m_{F}=+2$ ground-state Zeeman sublevels.
The two-photon detuning was varied by slowly sweeping the longitudinal
magnetic field near zero.
The observed EIT resonance in Fig.~\ref{tests.fig}b had a full width
$\approx 50$ Hz (coherence lifetime $\approx 6.6$~ms).


In both double-resonance (DR) and EIT measurements, the width of the
narrow resonance is limited by the decoherence rate of the Zeeman
sublevels.
To set a limit on the contribution of wall collisions,
we work at low laser and rf field power to avoid power broadening. We
also avoid broadening from Rb-Rb spin-exchange collisions by keeping
the cell at a low temperature ($36\:{}^\circ$C, corresponding to a Rb
number density of $3\times 10^{10}~\mathrm{cm}^{-3}$).  We believe
that uncompensated magnetic field gradients are the leading remaining
decoherence source in both DR and EIT linewidth measurements,
which is consistent with the approximately factor of two difference in
the measured DR and EIT linewidths:
for a fixed field gradient, the $\Delta m = 2$ EIT transition should have
twice the frequency width of the $\Delta m =1$ double resonance
measurement.
These results imply that the contribution of wall collisions to the Rb
Zeeman decoherence rate is $\ll10$ Hz.

\section{Dual-structure EIT and slow light measurements
  \label{slowlight}}

%
%

\begin{figure*}
\includegraphics[width=1.6\columnwidth]{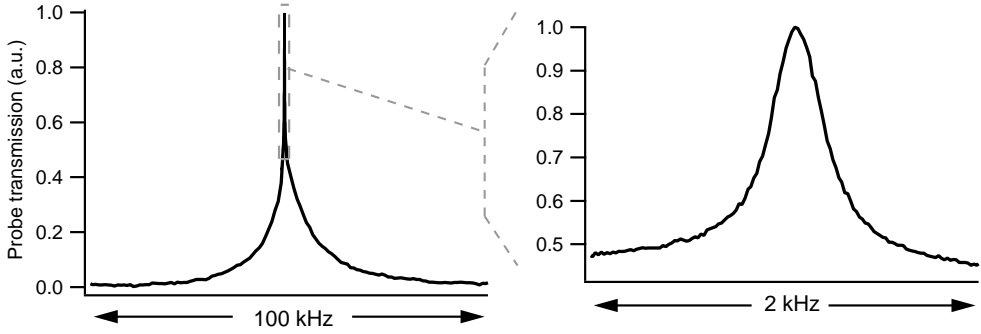}
\caption{Measured dual-structure EIT lineshape characteristic of a coated
  cell.  The full widths are $13$ kHz for the broad structure and
  $350$ Hz for the narrow peak.  Note that the narrow peak is
  substantially narrower than the width of $\sim10$ kHz expected from
  transit time broadening across the 4.5 mm beam.  Control field
  intensity is $3.5$~mW/cm$^2$ and cell temperature is
  $48\:{}^\circ$C; hence the narrow EIT peak is subject to moderate
  power broadening. }
\label{dualEIT.fig}
\end{figure*}

Figure \ref{dualEIT.fig} shows an example of the measured
dual-structure EIT lineshape in a paraffin-coated Rb vapor cell.
The EIT lineshape consists of two distinct features: a broad pedestal
due to atoms interacting with the laser beam only once, and a narrow
central peak due to atoms returning to the beam after multiple bounces
from the walls (and thus having a much longer coherence lifetime).
A comprehensive study of EIT lineshapes and slow light in coated cells
will be presented in future publications.

%

\begin{figure*}
\includegraphics[width=1.4\columnwidth]{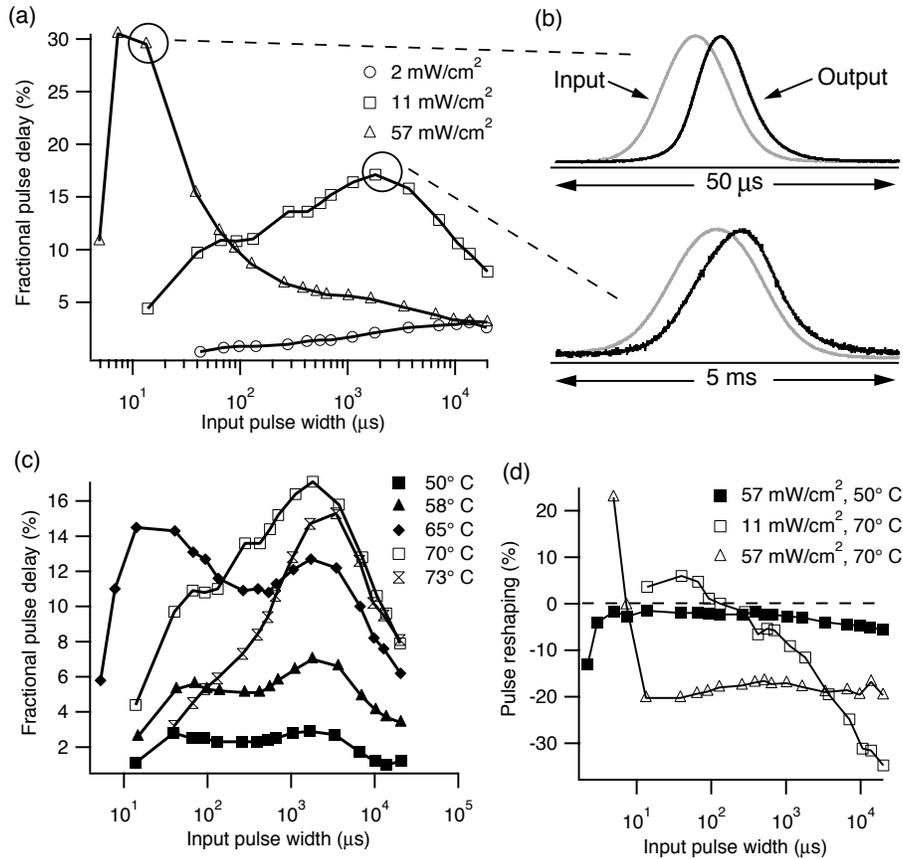}
\caption{Fractional pulse delay as a function of input pulse width
  for (a) different laser intensities at constant cell temperature
  ($70\:{}^\circ$C) and (c) different cell temperatures at constant
  laser intensity ($11$~mW/cm$^2$). (b) Examples of slow light pulse
  propagation for two maximum fractional delay regimes. (d) Slow light
  pulse reshaping (\emph{see text}). Beam diameter $\approx2$ mm.
} \label{dualFracDelay.fig}
\end{figure*}

We explored slow light pulse propagation over a wide range of
experimental parameters (field intensities between $1$ and $60$
mW/cm${}^2$, cell temperatures from $50$ to $75\:{}^\circ$C, and pulse
widths from 1 $\mu$s to 20 ms). See Fig.~\ref{dualFracDelay.fig}.
We found two distinct regimes of maximum fractional pulse delay,
corresponding to the two frequency scales observed in the
dual-structure EIT lineshapes.  At high laser intensity, many atoms
are pumped into the dark state on a time scale comparable to the
transit-time of an atom through the laser beam (as in an uncoated
cell). This leads to an observed maximum fractional delay for pulse
lengths near 10 $\mu$s (triangles in Fig.~\ref{dualFracDelay.fig}a),
corresponding to the broad EIT pedestal in Fig.~\ref{dualEIT.fig}. At
lower laser powers (squares in Fig.~\ref{dualFracDelay.fig}a), the
maximal fractional delay is observed for much longer pulses with
lengths of several milliseconds. These pulses match the narrow central
feature of the EIT resonance of Fig.~\ref{dualEIT.fig}.
%
%

Naively, we expect broadening of output pulses (\emph{i.e.}, narrowing
of the frequency spectrum) when the pulse bandwidth is larger than the
EIT spectral window~\cite{boyd05}. However, in these measurements, we
typically observe negative fractional pulse reshaping (defined as the
difference of the input and output pulse widths, normalized to the
input pulse width) as shown in Fig.~\ref{dualFracDelay.fig}d. We are
currently studying this counterintuitive phenomenon in greater detail.

\begin{figure*}
\begin{center}
\includegraphics[width=1.6\columnwidth]{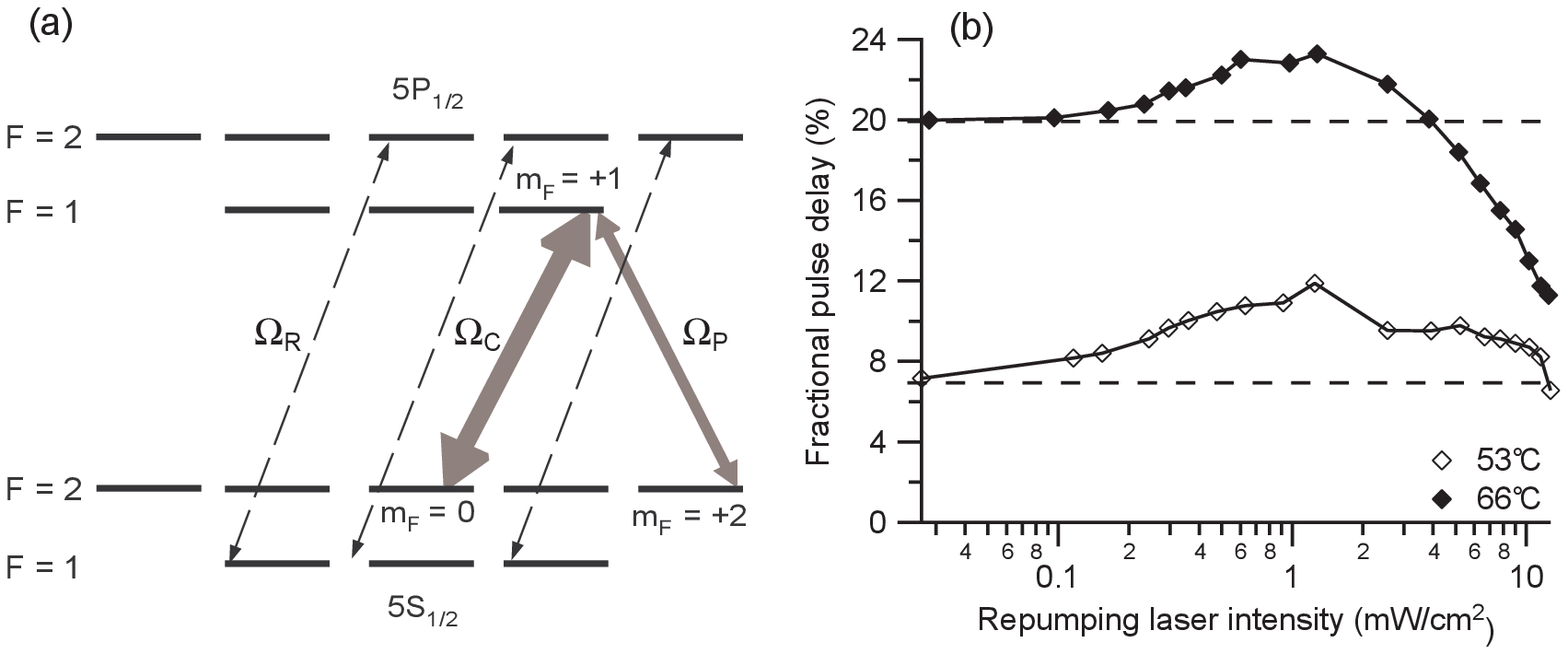}
\end{center}
\caption{ (a) ${}^{87}$Rb level scheme showing the
  repumper ($\Omega_R$) and the fields used in EIT and slow light
  experiments. (b) Slow light fractional delay vs.\   power of
  repumping laser. Dashed horizontal lines show the fractional delay
  without repumping laser. }
\label{FIGrepumper}
\end{figure*}

By adding a repumping beam on the lower energy hyperfine level, we can
increase the number of atoms interacting with the laser fields without
increasing the atomic density.  The repumper ($\Omega_R$ in
Fig.~\ref{FIGrepumper}a) depopulates the $F=1$ hyperfine ground state
using a laser field resonant with the $F=1\rightarrow{}F^\prime=2$
transition in ${}^{87}$Rb and circularly polarized with the same
polarization as the control field.  At optimized repumping intensity,
the fractional delay of a slow light pulse is increased
(Fig.~\ref{FIGrepumper}b); the fractional delay nearly doubles for low
densities and therefore relatively low fractional delays (open
diamonds in Fig.~\ref{FIGrepumper}b). At higher densities and delays,
however, the benefit of the repumping beam is less significant (solid
diamonds in Fig.~\ref{FIGrepumper}b).

\begin{figure*}
\begin{center}
\includegraphics[width=1.6\columnwidth]{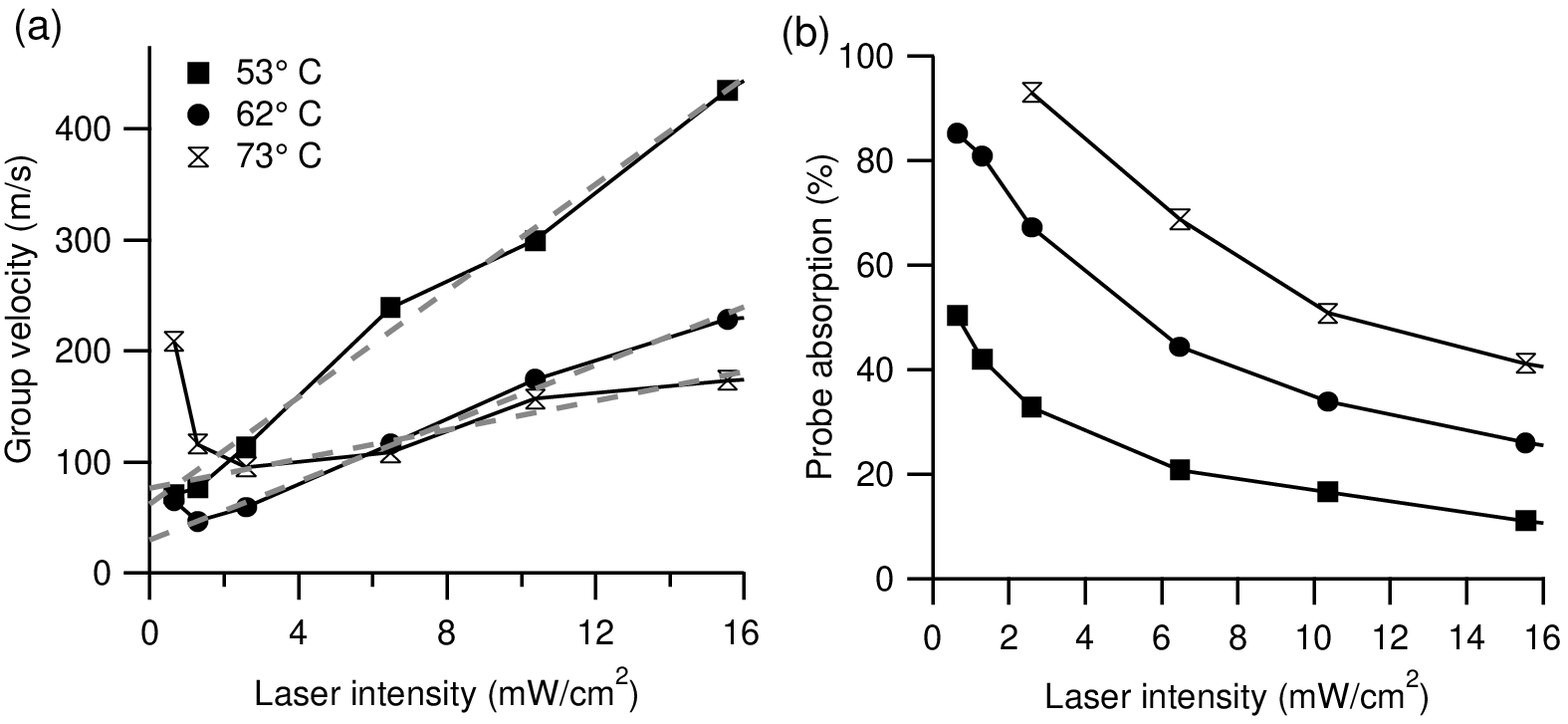}
\end{center}
\caption{Dependence of (a) group velocity and (b) pulse absorption
  on control field intensity for different cell temperatures. Dashed
  lines indicate linear trending with control field intensity expected
  without radiation trapping. } \label{FIGradiationtrapping}
\end{figure*}

The maximum fractional pulse delays observed in these coated cells do
not exceed $30\%$.
We believe that our observed delays are limited by radiation trapping,
\emph{i.e.,} the re-absorption of spontaneous, incoherent photons by
the atomic medium~\cite{radtrap}. Although fluorescence is suppressed
by EIT, spontaneously emitted photons are still present due to
residual absorption such as that from other excited state
levels~\cite{matsko05}.
If an atom in the dark state absorbs an unpolarized photon, its
coherence is destroyed. Thus radiation trapping effectively increases
the ground state decoherence rate, leading to higher absorption,
broader EIT linewidth, and shorter pulse delays.
Motional averaging of atoms with ground state coherence throughout the
vapor cell exacerbates the effects of radiation trapping in coated
cells in comparison to buffer gas cells in which atoms participating
in the slow light process are typically close to the axis of the cell.

For long pulses (in which the pulse bandwidth is much narrower than
the width of the narrow, central EIT feature) we expect the group
velocity to be proportional to the control field intensity and
inversely proportional to the atomic
density~\cite{matsko_slreview,boyd_slreview}. At low atomic densities
for which most spontaneous photons are not reabsorbed (squares in
Fig.~\ref{FIGradiationtrapping}a) we observe such linear scaling of
the group velocity with laser intensity.  Similarly, at relatively
high laser intensities, for which the additional decoherence due to
radiation trapping is insignificant compared to the power-broadened
EIT bandwidth, the measured group velocity scales inversely with
atomic density (Fig.~\ref{FIGradiationtrapping}a). However, at low
control field intensities and high densities the group velocity begins
to increase (Fig.~\ref{FIGradiationtrapping}a) as the absorption of
the probe beam by the slow light medium becomes substantial
(Fig.~\ref{FIGradiationtrapping}b).  This observation also explains
why improved state selection from the repumping laser has only a
limited effect at higher temperature: an increase in the density of
appropriately state-selected atoms further enhances absorption of the
probe field.

\section{Conclusion}

We have studied EIT and slow light pulse propagation in
paraffin-coated Rb vapor cells, and outlined the methods used in their
manufacturing and testing.  The long lifetimes for ground-state spin
coherence due to the wall coating lead to a narrow central peak in the
EIT lineshape and hence ultra-slow group velocity pulse propagation,
which in principle should enable very long light delay and storage
times.

To properly describe pulse propagation in a coated cell, both atoms
interacting multiple times with the beam after repeated wall
collisions and
atoms interacting only once  with the laser beam must be included.
These two interaction timescales correspond to a dual-structure EIT
lineshape with two regimes for fractional pulse delay optimization.
%
In our studies to date we believe that radiation trapping is the
leading factor limiting observed slow-light fractional delays to
$\approx30\%$. To achieve larger fractional delays in coated cells, we
are presently pursuing long, narrow cell geometries and the use of
isotopically enriched ${}^{87}$Rb to reduce the role of radiation
trapping.

The authors are grateful to M.~A. Hohensee and Y. Xiao for useful
discussions. This work was supported by DARPA, ONR and the Smithsonian
Institution.


\end{document}